\documentclass[]{spie}  

\usepackage[]{graphicx}
\usepackage[T1]{fontenc}
\usepackage{pslatex}

\title{PANIC: A Near-infrared Camera for the Magellan Telescopes }

\author{Paul Martini\supit{a,b}, S.~E. Persson\supit{a}, D.~C. Murphy\supit{a}, C. Birk\supit{a}, S.~A. Shectman\supit{a}, \\ 
S.~M. Gunnels\supit{c}, and E. Koch\supit{d}
\skiplinehalf
\supit{a}Observatories of the Carnegie Institution of Washington, 813 Santa Barbara Street, Pasadena, CA 91101, USA; \\
\supit{b}Harvard-Smithsonian Center for Astrophysics, 60 Garden Street, MS20; Cambridge, MA 02138, USA; \\
\supit{c}Paragon Engineering, 18231 Old Ranch Road, Tehachapi, CA 93561, USA; \\
\supit{d}Dedicated Micro Systems, 632 South Park Rose Avenue, Monrovia, 
CA 91016, USA \\
}

\authorinfo{Further author information: (Send correspondence to P.M., S.E.P.)\\P.M.: E-mail: pmartini@cfa.harvard.edu, Telephone: 1 617 496 1809\\  S.E.P.: E-mail: persson@ociw.edu, Telephone: 1 626 304 0235}

\newcommand{\aj}{Astronom. Journal}
 
\newcommand{\pasp}{Pub. Astro. Soc. Pac.} 
 
\begin{document} 
  \maketitle 

\begin{abstract}

PANIC (Persson's Auxiliary Nasmyth Infrared Camera) is a near-infrared 
camera designed to operate at any one of the $f$/11 folded ports of the 6.5m 
Magellan telescopes at Las Campanas Observatory, Chile. 
The instrument is built around a simple, all-refractive design 
that reimages the Magellan focal plane to a plate scale of 0.125$''$ 
pixel$^{-1}$ onto a Rockwell 1024x1024 HgCdTe detector. 
The design goals for PANIC included excellent image quality to sample 
the superb seeing measured with the Magellan telescopes, high throughput, a 
relatively short construction time, and low cost. 
PANIC has now been in regular operation for over one year and has proved to 
be highly reliable and produce excellent images. The best recorded image
quality has been $\sim 0.2''$ FWHM. 

\end{abstract}



\section{Instrument Overview}
\label{sec:intro}  

Considerable interest in the rapid development of a near-infrared (NIR) camera 
for the Magellan Project was generated after first light 
for the 6.5m Magellan~I (Walter Baade) telescope in September 2000. 
The main requirements included excellent image quality 
to sample the superb seeing observed at Magellan, high-throughput, and 
a simple design that could be completed relatively quickly. 
In addition, the instrument was envisioned to eventually occupy one of the 
folded ports (formerly known as auxiliary Nasmyth ports) and thus remain 
available for use at all times. 

The ready availability of the camera during any observing run was motivated 
by one of the science goals: time-critical observations of temporally variable 
objects, such as supernovae and gamma ray bursts. 
Scientific interests of astronomers at the Carnegie Observatories and 
other members of the Magellan consortium included follow-up observations of 
interesting objects discovered in wide-field infrared surveys on smaller 
telescopes, high-redshift quasars and clusters of galaxies, counterparts 
to Galactic X-ray sources, and searches for the galaxies responsible 
for quasar absorption systems. 

These science goals led to PANIC (Persson's Auxiliary Nasmyth Infrared Camera), 
an all-refractive, $1 - 2.3$ $\mu$m camera that reimages the $f/11$ input beam 
of the 6.5m Magellan telescopes to produce a plate scale of 
$0.125''$ pixel$^{-1}$ on its Rockwell HAWAII HgCdTe focal plane 
array (1024x1024 18.5 $\mu$m pixels). 
Use of a compact, all-refractive design and adoption of the electronics and 
control software from the recently completed wide-field, NIR camera 
WIRC\cite{persson02} minimized development costs and led to a 
relatively short construction time. 

\section{Optics} 
\label{sec:optics}

The optical design was optimized to critically sample the best images 
produced by the Magellan telescopes. Typically, these telescopes 
produce images of $0.5''$ or better at visible-wavelengths and a plate 
scale of $0.125''$ pixel$^{-1}$ was chosen to sample the expected 
image quality at near-infrared wavelengths. The principal and folded ports 
of the Magellan telescopes are designed to be fed by the Gregorian $f/11$ 
secondary mirror, which produces a focal plane scale of $0.347$mm per 
arcsecond. 
PANIC reimages the telescope focal plane from $f/11$ to $f/4.7$ and produces 
a $128'' \times 128''$ field of view.

Two other key considerations for the optical design were high-throughput in a 
compact design and a relatively warm temperature for the camera optics 
support structure of 200 K (compared to 77 K more typical of NIR instruments). 
The significant cost savings and ease of construction associated with a smaller 
instrument motivated an all-refractive design. The glasses used include 
IR-grade fused silica, CaF2, PbH1, and the O'Hara glass S-FTM16, where the 
latter was chosen because it forms an excellent infrared achromat with CaF2. 
IR-grade fused silica was chosen for the vacuum window because of its excellent 
durability. One key advantage of the warmer optical elements lies in 
reduced uncertainty in the refractive indices, particularly S-FTM16. 
The refractive indices for S-FTM16 and PbH1 at 200~K and 77~K, respectively, 
were calculated with a temperature-dependent refractive index 
model\cite{tropf95}. This model adds temperature-dependent terms to the 
common Sellmeier dispersion formula and successfully reproduces the 
measured temperature-dependent dispersion for CaF2, among other materials. 
There are also mechanical and thermal advantages discussed below in 
sections~\ref{sec:mech} and \ref{sec:thermal}. 

\begin{figure}[!ht]
\begin{center}
\begin{tabular}{c}
\includegraphics[width=16.0cm]{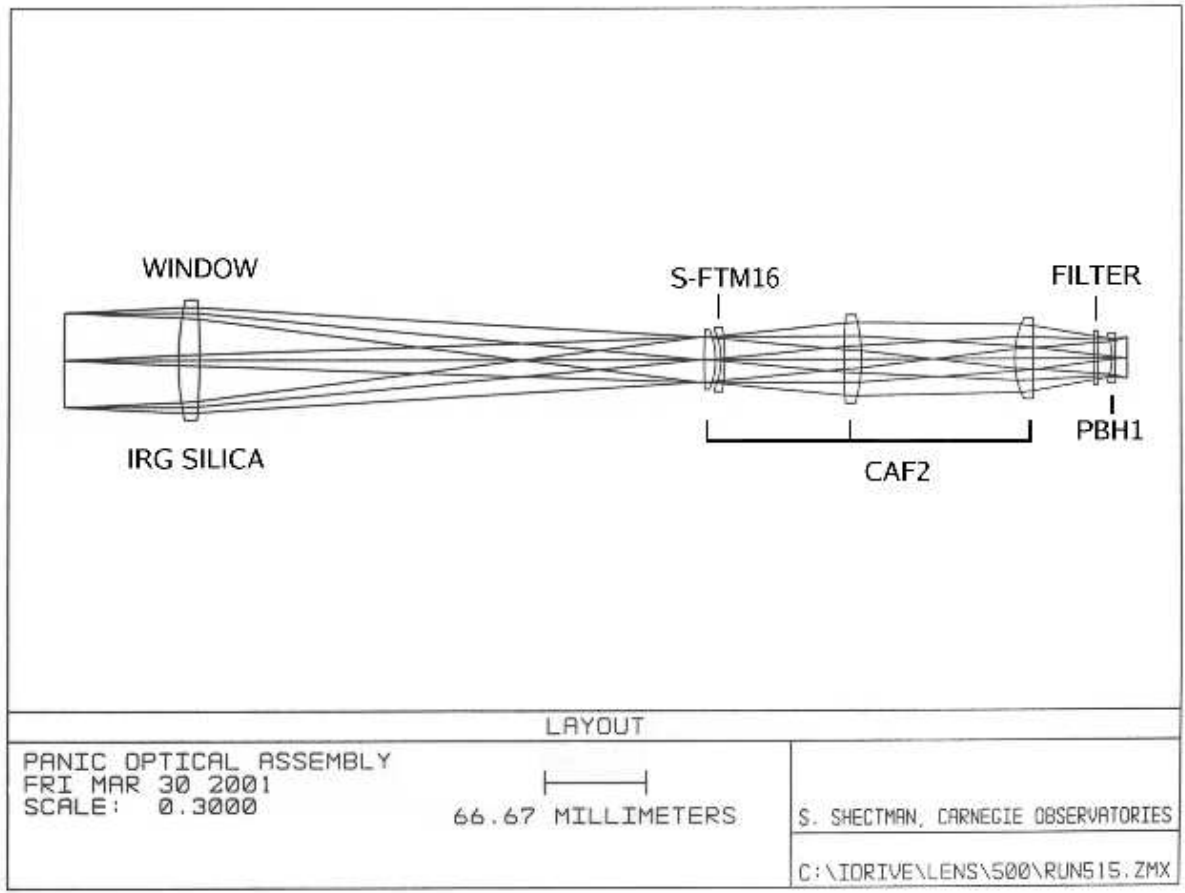} 
\end{tabular}
\end{center}
\caption[Optical design]
{ \label{fig:optics}
Optical design. The first element on the left is the vacuum window of 
IR-grade fused silica. 
The central four elements are housed in the ISF as described in 
Section~\ref{sec:mech}, while the PbH1 field flattener is incorporated into 
the detector mount. 
The flat element immediately prior to the field flattener marks the 
approximate location of the filter wheel. 
}
\end{figure}

The optical design is listed in Table~\ref{tbl:optics} and shown in 
Figure~\ref{fig:optics}. The all-spherical optical elements were manufactured 
by Janos Technology Inc. 
The design produces an rms spot radius of 3 $\mu$m on axis and better than 
4 $\mu$m images at the edges of the field. 
Because the field lens, which serves as the vacuum window of the dewar, is not 
an achromat the position of the pupil image is mildly wavelength dependent.  
Since the cold stop is most critical for masking thermal radiation from the 
telescope structure surrounding the secondary mirror, the fixed cold stop in 
PANIC was positioned and sized for the red end of the NIR $K_s$ filter 
at approximately 2.3 $\mu$m, the wavelength region where thermal radiation 
is most significant.
Another trade off in this simple design is the lack of a 
well-collimated beam, which precludes addition of a grism for spectroscopy. 

\begin{table}[h]
\caption{PANIC optical prescription. All dimensions are in mm.} 
\label{tbl:optics}
\begin{center}
\begin{tabular}{|l|l|l|l|l|l|} 
\hline
\rule[-1ex]{0pt}{3.5ex} Element & Surface & Radius & Thickness & Material & Diameter \\
\hline
\rule[-1ex]{0pt}{3.5ex} PAN1    & 1       &  192.35   & 14.    & IR-grade Fused Silica   & 80 \\
\hline
\rule[-1ex]{0pt}{3.5ex}         & 2       & -551.55   & 336.5  &          &    \\
\hline
\rule[-1ex]{0pt}{3.5ex} PAN2    & 3       &  414.16   & 7.514  & CaF2     & 40 \\
\hline
\rule[-1ex]{0pt}{3.5ex}         & 4       & -50.82    & 3.5    &          &    \\
\hline
\rule[-1ex]{0pt}{3.5ex} PAN3    & 5       & -48.248   & 2.5022 & S-FTM16  & 43 \\
\hline
\rule[-1ex]{0pt}{3.5ex}         & 6       & -137.83   & 69.7   &          &    \\
\hline
\rule[-1ex]{0pt}{3.5ex} PAN4    & 7       &  437.512  & 11.521 & CaF2     & 59 \\
\hline
\rule[-1ex]{0pt}{3.5ex}         & 8       & -115.39   & 111.4  &          &    \\
\hline
\rule[-1ex]{0pt}{3.5ex} PAN5    & 9       &  64.2     & 12.022 & CaF2     & 54 \\
\hline
\rule[-1ex]{0pt}{3.5ex}         & 10      & -4044.36  & 54.130 &          &    \\
\hline
\rule[-1ex]{0pt}{3.5ex} PAN6    & 11      & -42.333   & 2.004  & PbH1     & 33 \\
\hline
\rule[-1ex]{0pt}{3.5ex}         & 12      &  321.741  & 8.     &          &    \\
\hline
\end{tabular}
\end{center}
\end{table}

\section{Mechanical Design} \label{sec:mech}

\begin{figure}[!ht]
\begin{center}
\begin{tabular}{c}
\includegraphics[width=16.0cm]{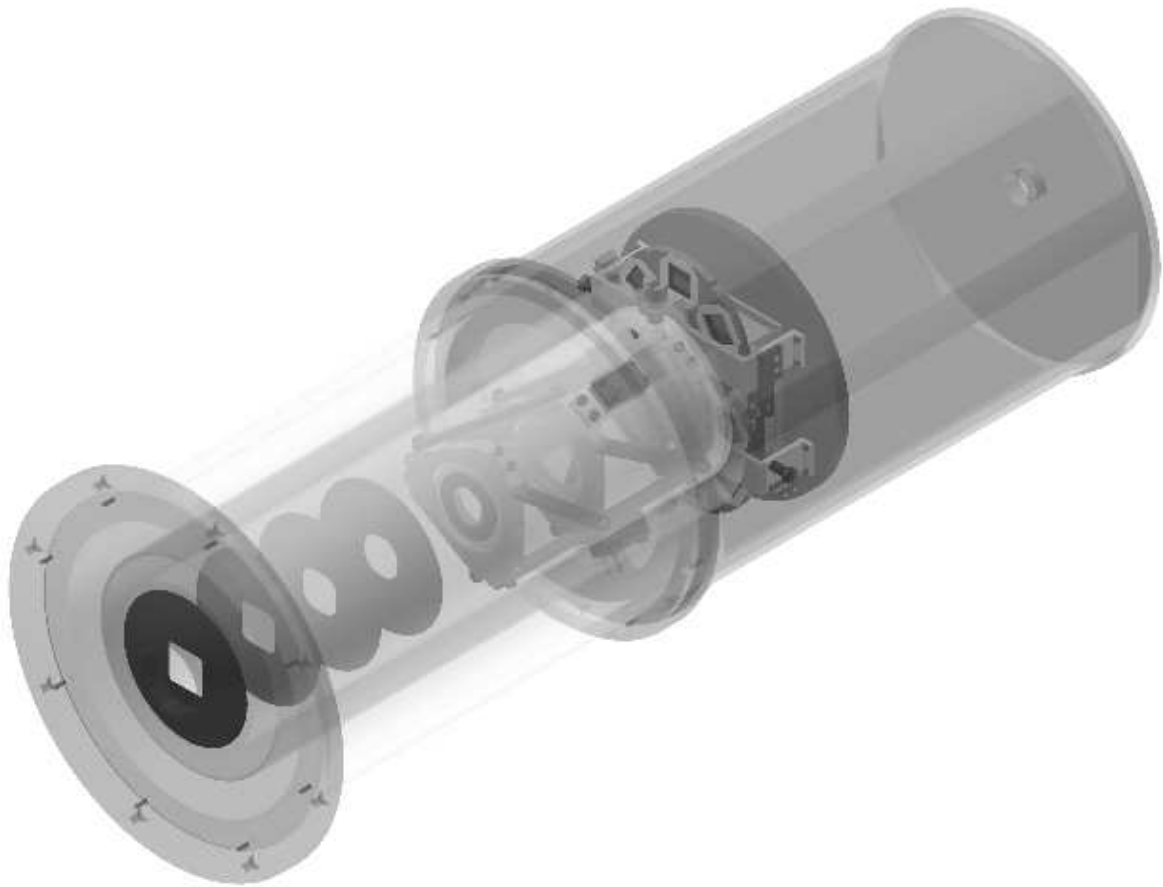}
\end{tabular}
\end{center}
\caption[Mechanical design]
{ \label{fig:mech}
Mechanical design. The main sections shown are the optics housing 
(left) and the ND-10 dewar from Infrared Laboratories (right). The optics 
housing includes a radiation shield and baffles and is encased in an 
outer aluminum tube (not shown). The detector and filter 
wheels are mounted directly to the LN$_2$ dewar. The ring between these 
two main sections supports the ISF, which holds lenses two through five. 
}
\end{figure}

The mechanical design is centered around support for the optical elements and 
the requirement to minimize flexure, which would produce element tilt and 
decenter during rotation about the optical axis. 
Elements 2--5 are supported in a space frame built of Invar, which was chosen 
for its low thermal expansion properties. 
This Invar space frame (ISF) is mounted in a stainless steel ring. 
This ring is rigidly 
supported by an aluminum tube that connects it to the front mounting surface 
and also serves as the vacuum enclosure. The mechanical layout is 
illustrated in Figure~\ref{fig:mech} (note that the outer aluminum tube 
is not shown). 
Finite element analysis demonstrates less than 2 $\mu$m lateral deflection 
between the focal plane and ISF at Nasmyth, less than 1 $\mu$m between 
the field lens and ISF, and less than $1'$ tilt. 
These deflections and tilts are well within the allowances of the optical 
design. 
The liquid nitrogen (LN$_2$) reservoir is a model ND-10 dewar from Infrared Laboratories and is 
also bolted to the stainless steel ring. 
This dewar has a capacity of 15 liters. 

\subsection{Optical Mounts} \label{sec:mounts}

The mounts for optical elements two through five are supported by the ISF 
described 
above. The ISF includes one Invar support frame for the closely-spaced elements 
two and three, and additional, separate supports for elements four and five. 
These 
support frames hold separate lens mounts for each of the elements, where the 
lens mounts are built from materials chosen to have comparable thermal 
expansion properties to the lenses. For the CaF2 elements, the aluminum alloy 
206.0-T7 provides a good match, while titanium provides a good match to 
S-FTM16. 
A thin layer of paper lines each lens mount in order to provide an additional 
buffer against thermal and mechanical stresses. 
Thermal contact points between the lens mounts and the ISF were also minimized 
to insure that the mounts and the lenses cool radiatively, which further 
alleviates stress on the optical elements. The mount for element six, 
which serves as a field flattener, is incorporated directly into the 
detector mount. 

\subsection{Filter Wheels} \label{sec:filters}

There are two, six-position filter wheels in PANIC. One of these 
wheels is populated with $Y$, $J$, $H$, and $K_s$ broadband filters; 
the remaining two are open and dark positions. The second wheel contains 
two 1\% narrowband filters centered at 2.12 $\mu$m ($H_2$ 1-0 S(1) 
line) and 2.16 $\mu$m (Br$\gamma$), although their presence in a 
converging beam broadens the passband. 
They are nevertheless useful for morphological studies of emission-line 
regions. 
A series of medium-band filters are planned for the remaining open 
wheel positions. 
The filter wheel mechanisms are bolted directly to the work surface of the 
dewar immediately above the detector mount and field flattener. The 
wheels cool conductively to nearly 77K. 

The filter wheel positions are changed by external stepper motors, which are 
coupled through the dewar walls with feedthroughs from Ferrofluidics 
(now Ferrotec). These feedthroughs are thermally isolated 
from the wheels with G-10 insulators and additional radiation 
shields. The internal gears for the wheels are mirrored outside the 
dewar and drive an encoder barrel, which is used to provide an absolute 
reference for each filter position. Two microswitches inside the dewar 
are used to determine if the wheels are in one of the spring-loaded 
detent positions. These detents insure that the filter positions are 
exactly reproducible. 

\section{Thermal} 
\label{sec:thermal}

One important simplification of the PANIC design was the decision to limit 
the upper wavelength range of the instrument to the red end of the $K_s$ 
filter at 2.32 $\mu$m, a decision which significantly reduces the thermal 
requirements for the instrument. Calculation of the relative thermal emission 
from the window and the remainder of the instrument showed that at an 
internal temperature of 200 K the instrument was still an insignificant 
contributor to the background through the $K_s$ filter. This operating 
temperature of 200 K, rather than near the LN$_2$ temperature of 77 K, results 
in the less stringent optical and mechanical requirements discussed 
previously, as well as more rapid thermal cycling of the instrument. 

However, the decision to maintain most of the instrument at 200 K did 
necessitate careful attention to thermal stability to both insure that the 
optics would remain at their designated temperature and to minimize any 
variations in the instrumental background. Thermal stability was achieved in 
two steps using both passive and active methods: First, the thermal and 
conductive loads on the radiation shield and ISF were calculated and tuned 
to produce an equilibrium temperature of approximately 200 K. 
These temperatures were coarsely tuned by wrapping the outside of the 
radiation shield with multiple layers of aluminized Mylar and applying 
space-qualified black paint to the optical baffles and ISF. 
The temperatures were then adjusted more finely with two heat switches 
incorporated into the dewar by Infrared Laboratories. The heat switches 
were used to separately regulate the degree of thermal conductivity between 
the LN$_2$ reservoir and both the radiation shield and ISF. 
A LakeShore 321 Autotuning Temperature Controller is then used to actively 
maintain the temperature of the ISF at precisely 200 K through a feedback 
loop between the heater and a temperature sensor, both located on the exterior 
of the ISF. 

We employ a LakeShore 218 Temperature Monitor to continuously monitor 
eight temperature sensors located on the detector mount, ISF, 
and radiation shield. The temperature sensor on the detector mount is also 
used to monitor the cooldown rate of the detector relative to Rockwell's 
requirement of less than 1 K per minute. Because of this requirement, the 
cooldown process for the detector requires several hours. Approximately 
24 hours are required for the ISF and lenses to come into equilibrium. 
The LN$_2$ reservoir has a 15 liter capacity and is filled halfway due to 
its horizontal orientation (at Nasmyth) and rotation of the instrument 
about the optical axis. LN$_2$ is added to PANIC daily and it is maintained 
at cryogenic temperatures for rapid availability on any night. 

\begin{figure}[!ht]
\begin{center}
\begin{tabular}{c}
\includegraphics[width=16.0cm]{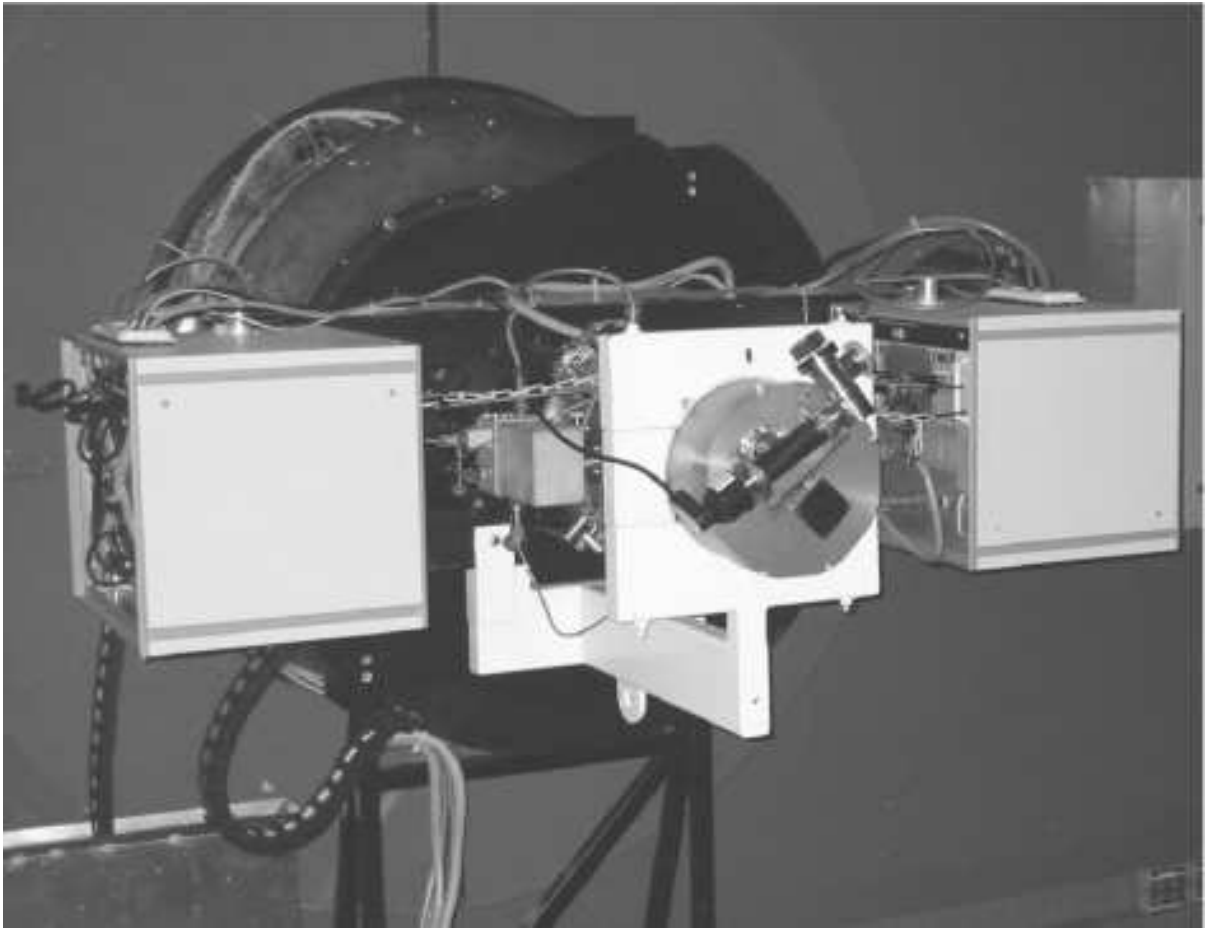}
\end{tabular}
\end{center}
\caption[PANIC]  
{ \label{fig:panic}
PANIC mounted on the East Nasmyth instrument rotator of the 6.5m Magellan~I 
(Walter Baade) telescope in April 2004. The two, small electronics racks 
described in Section~\ref{sec:elect} are mounted on either side of PANIC 
and rotate with the instrument. The central square structure ({\it white}) 
around the dewar is a mounting fixture, while a freestanding guide for the 
cable carrier ({\it black}) is also visible. 
}
\end{figure}

\section{Electronics and Control System} 
\label{sec:elect}

The HAWAII detector in PANIC was chosen because it offers 
good quantum efficiency, was readily available, affordable, and we had 
considerable experience with this device. We read the array using nearly 
identical electronics to those used for the WIRC camera\cite{persson02} at the 
2.5m du Pont telescope, as well as previous generations of NIR 
instruments built at the Carnegie Observatories\cite{murphy95}. 
Briefly, all four quadrants of the array are read out in parallel through 
external, low-noise amplifiers. These amplifiers remove the DC bias voltage and 
amplify the pixel voltage by a factor of ten. The resultant signals from each 
quadrant and amplifier are then read through their own dedicated 16 bit 
analog-to-digital converter (ADC). The output from the ADCs are latched 
into a 16 bit PCI parallel interface card (from Spectral Instruments) 
in the data acquisition computer. 

Two small electronics racks are mounted directly to the instrument rotator 
on either side of PANIC (see Figure~\ref{fig:panic}). 
These racks include low-noise power supplies, the waveform generator, the 
ADC unit, a Telebyte 273 Fiber Optics Mux for the 
clocking signals, a Cold Cathode Model 943 pressure gauge from MKS 
Instruments, and temperature sensors for these racks. 
The data acquisition computer was custom built from readily available 
components, including a 1.6GHz Athlon processor, and runs Windows 98 SE. 
It is located in an external instrumentation rack, which is currently located 
on the Nasmyth platform near the instrument. A cable carrier from Igus Inc.\ 
is used to support the cables that run between the rotating and nonrotating 
components. 
The data acquisition computer commands the waveform generator to generate the 
clocking signals for array control. These commands are sent via the Fiber 
Optics Mux. 
The external rack also contains the two LakeShore units described in the 
previous section, the filter wheel motor controllers, the ADC power supply, 
rack temperature sensors, and an Equinox Ethernet Serial Provider, which 
transfers signals from 
the instrument control computer to the filter wheel motor controller and from 
the LakeShore units to the control computer. 
The low-noise amplifiers described previously are mounted directly to the 
dewar, as are the filter wheel motors and encoders. 

The external rack is connected via Gigabit Ethernet to the instrument 
control computer, which has a 1.6GHz Athlon processor and runs the Windows 
2000 operating system, and a data analysis computer, which has a 
2.4GHz Athlon processor and runs the RedHat Linux operating system. 
The instrument control computer has a series of GUIs that include a range 
of preset observation macros and readouts for the LakeShore temperature 
units, in addition to standard control of exposure sequences and the filter 
wheels. The data analysis workstation includes an array of five 18Gb Seagate 
15,000 rpm SCSI drives for rapid data handling and standard 
astronomical data analysis software. Images are written to the data analysis 
workstation using network folders and the Samba server software. 
Both of these computers are currently accessed over the local network by 
observers using a dual-monitor workstation in the telescope control room. 
The instrument control computer is accessed remotely with the VNC Viewer 
software by RealVNC, while the data analysis computer is accessed via a 
standard secure shell connection. 

To supplement the standard data analysis tools in IRAF, we have developed 
a specialized IRAF package. This {\tt panic} package 
contains a set of tools that are resident on the data analysis workstation 
and can easily be downloaded and installed on an observer's personal 
workstation for further analysis after leaving the Observatory. The package 
includes a number of quick-look tools to difference images and perform 
quick sky subtraction as data are obtained, as well as a more complete data 
processing pipeline. 
In particular, these scripts take advantage of the excellent 
repeatability of the telescope motions to compensate for telescope offsets 
and allow accurate registration of the images so that single or multiple dither 
sequences can be combined automatically and without degradation of image 
quality. 
The {\tt panic} package is available from the instrument web 
site\footnote{http://www.ociw.edu/lco/magellan/instruments/PANIC/panic/index.html}, 
which also contains complete documentation for observers. 

\section{Commissioning and Performance} 
\label{sec:perform}

\begin{figure}[!ht] 
\begin{center}
\begin{tabular}{c}
\includegraphics[width=8cm]{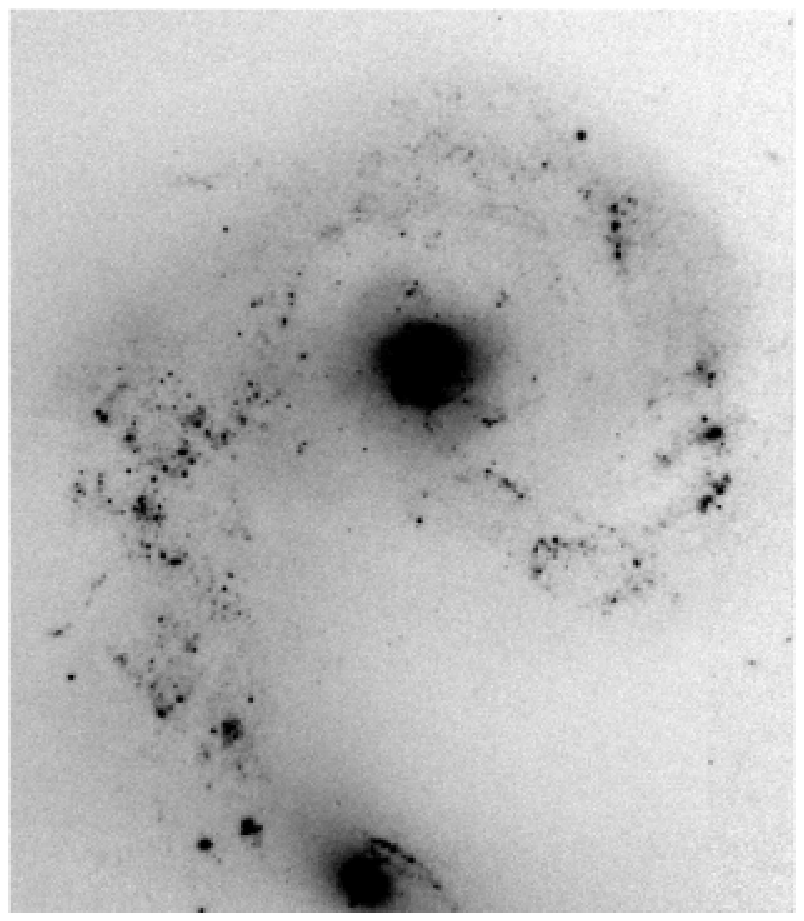}
\includegraphics[width=8cm]{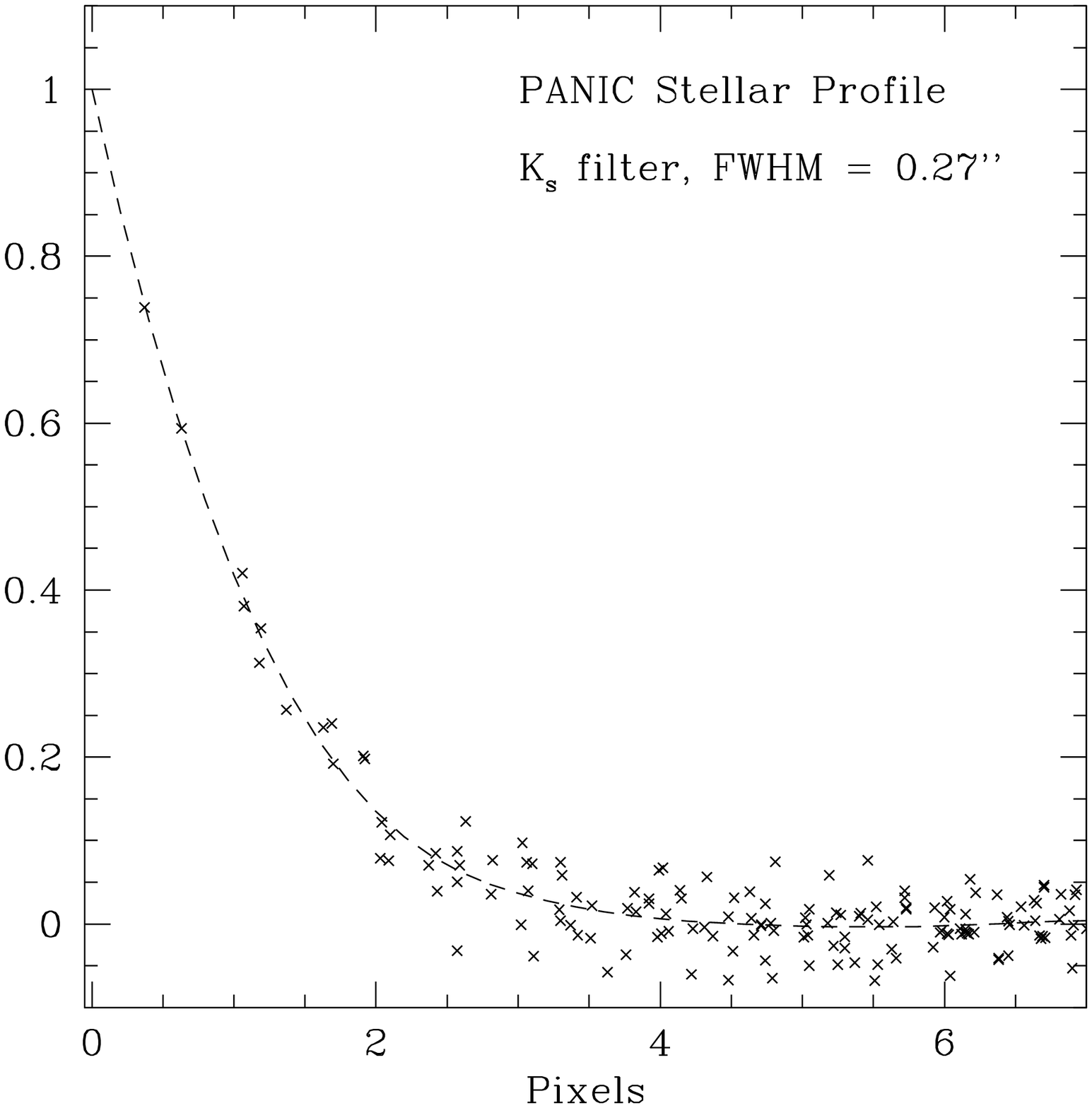}
\end{tabular}
\end{center}
\caption[Antennae]  
{ \label{fig:antennae}
({\it left}) The Antennae Galaxy (NGC4038/9) observed with PANIC on 
April 11, 2003 through the $K_s$ filter. 
This is the sum of multiple exposures for a total integration time of 750s.  
The average FWHM of stellar sources in this image is $0.27''$ and the 
field of view is $100'' \times 115''$. 
({\it right}) Radial profile of an unresolved source in the Antennae image. 
A fit to the data is also shown. 
}
\end{figure}

PANIC was successfully commissioned on the Magellan~II (Clay) telescope in 
April 2003 and immediately went into regular operation. Aside from 
standard instrument checkout and calibration frames, the main task 
for the commissioning run was accurate alignment of the instrument pupil. 
Because PANIC has a fixed pupil mask, 
the pupil alignment was performed by tip-tilting the instrument 
on the rotator mount. This was accomplished by identifying a minimum in 
the sky background through the $K_s$ filter while pointing at a region of 
relatively blank sky. 

The standard mode of detector readout is through the common technique of 
double-correlated sampling. The detector gain, readnoise, and linearity were 
measured and found to be well within expected values. The gain is approximately 
$2.4\ e^-$ DN$^{-1}$ and the readnoise has been reduced to nearly $15\ e^-$ 
after some additional grounding was performed at the telescope. 
The detector remains linear to better than 1\% below $36,000\ e^-$ at typical 
background rates (appropriate to the sky background flux through the $K_s$ 
filter on a warm night), deteriorating to 5\% nonlinear at $72,000\ e^-$.
Saturation occurs at approximately $108,000\ e^-$ counts. The detector nonlinearity is 
quite repeatable and corrected as part of the standard data processing 
pipeline described above. The dark current was measured and found to be 
less than $0.03\ e^-$ s$^{-1}$. Dark frames are not typically obtained, 
although they are sometimes used for cosmetic purposes when the sky 
counts do not dominate the noise statistics, or for twilight flats. 

During the commissioning period images as good as $\sim 0.2''$ FWHM were 
observed at $K_s$. The optical design is thus performing extremely 
well, even producing marginally undersampled data under the best 
conditions. More typical observed image quality is $0.3''$ FWHM. 
As an illustration of the image quality, a 12 minute $K_s$ exposure of 
the Antennae Galaxy is shown in Figure~\ref{fig:antennae}. Stellar sources 
are $0.27''$ FWHM. 

The standard mode of observations in the NIR is to obtain 
observations at many, slightly offset positions on the sky in order 
to obtain an accurate and precise measurement of the sky background. 
To maximize observation efficiency, a series of observation macros were 
incorporated into the data acquisition software that automatically 
make small offsets of the telescope and guide cameras. These 
observation macros include a number of five- and nine-position dither sequences 
and the size of the offset between each sky position can be tuned to 
account for the image quality or obtain a large area. The most efficient 
strategy is to maximize the time between readouts and between telescope 
offsets. For such cases the on-sky efficiency has been measured to 
approach 90\%. 

PANIC moved to the East Nasmyth port of the 6.5m Magellan~I (Walter Baade) 
telescope in July 2003 and is expected to move to one of the folded ports 
of the Baade telescope within the next year. During its first year of 
operation, PANIC has proved to be one of the most reliable instruments at the 
Magellan Observatory. It is currently in use on most bright nights
and routinely produces some of the best image quality, as is expected in the 
NIR. 

\acknowledgments

It is a pleasure to thank the machinists, technicians, and programmers 
who have contributed to PANIC: Estuardo Vasquez, Darrell Gilliam, Ken Clardy, 
Robert Storts, and Vincent Kowal. We also thank the excellent staff at 
Las Campanas Observatory in Chile, and particularly Miguel Roth and 
Oscar Duhalde for their assistance during commissioning. We also greatly 
appreciate the financial support provided by the Carnegie Observatories. 
PM acknowledges support from a Carnegie Starr Fellowship. 


\end{document}